\documentclass[12pt,a4paper]{article}
\usepackage{epsfig}
\usepackage{mathtools}
\usepackage{amsfonts}
\usepackage{amssymb}
\usepackage{amsmath}
\usepackage{color}
\usepackage{slashed}
\usepackage{cite}
\usepackage{caption}
\usepackage{subcaption}

\usepackage{tikz}
\usetikzlibrary{shapes}
\usetikzlibrary{trees}
\usetikzlibrary{calc}
\usetikzlibrary{decorations.pathmorphing}
\usetikzlibrary{decorations.markings}

\def\la              {\langle}
\def\ra              {\rangle}

\newcommand{\dotalpha}{{\dot{\alpha}}}

\newcommand{\abr}[1]{\langle #1 \rangle}

\newcommand{\vll}{{\smash{\lambda}}}
\newcommand{\vlt}{{\smash{\tilde{\lambda}}}}
\newcommand{\vlet}{{\smash{\tilde{\eta}}}}

\newcommand{\vlluu}{\smash{\underline{\underline{\smash{\lambda}}}}}
\newcommand{\vltuu}{\smash{\underline{\underline{\smash{\tilde{\lambda}}}}}}
\newcommand{\vleuu}{\smash{\underline{\underline{\smash{\tilde{\eta}}}}}}
\newcommand{\vletuu}{\smash{\underline{\underline{\smash{\eta}}}}}

\definecolor{grayn}{gray}{0.7}
\definecolor{lightgrayn}{gray}{0.8}

\newlength{\vacuumradius}
\newlength{\onshellradius}
\tikzstyle{db}=[circle, black, fill=black, minimum width=\onshellradius, draw, inner sep=0pt]
\tikzstyle{dw}=[circle, black, fill=white, minimum width=\onshellradius, draw, inner sep=0pt]
\tikzstyle{dvac}=[circle, black, fill=lightgrayn, minimum width=\vacuumradius, inner sep=0pt]
\tikzstyle{dl}=[circle, black, fill=white, inner sep=2pt]

\tikzset{
    gluon/.style={decorate, decoration={coil, amplitude=4pt, segment length=7pt, aspect=1}, draw=black}
}

\tikzset{->-/.style={decoration={
			markings,
			mark=at position .5 with {\arrow{>}}},postaction={decorate}}}

\newcommand{\C}{\mathbb{C}}
\newcommand{\T}{\mathbb{T}}

\newcommand{\PA}{\mathbb{PA}}

\newcommand{\bigslant}[2]{{\raisebox{.2em}{$#1$}\left/\raisebox{-.2em}{$#2$}\right.}}

\def\d{\text{d}}
\def\e{\text{e}}
\newcommand{\dbar}{\bar\partial}

\begin{document}

\begin{center}

\vspace{1cm}

{\bf \Large Ambitwistor strings and reggeon amplitudes \\ in $\mathcal {N}=4$ SYM.} \vspace{1cm}

{\large L.V. Bork$^{1,2}$ A.I. Onishchenko$^{3,4,5}$}\vspace{0.5cm}

{\it $^1$Institute for Theoretical and Experimental Physics, Moscow,
	Russia,\\
	$^2$The Center for Fundamental and Applied Research, All-Russia
	Research Institute of Automatics, Moscow, Russia, \\
	$^3$Bogoliubov Laboratory of Theoretical Physics, Joint
	Institute for Nuclear Research, Dubna, Russia, \\
	$^4$Moscow Institute of Physics and Technology (State University), Dolgoprudny, Russia, \\
	$^5$Skobeltsyn Institute of Nuclear Physics, Moscow State University, Moscow, Russia}\vspace{1cm}

\abstract{We consider the description of reggeon amplitudes (Wilson lines form factors) in $\mathcal {N}=4$ SYM within the framework of four dimensional ambitwistor string theory. The latter is used to derive scattering equations representation for reggeon amplitudes with multiple reggeized gluons present. It is shown, that corresponding tree-level string correlation function correctly reproduces previously obtained Grassmannian integral representation of reggeon amplitudes in $\mathcal {N}=4$ SYM. 
}
\end{center}

\begin{center}
\end{center}

\newpage



\section{Introduction}

The behavior of scattering amplitudes in high energy or Regge limit is determined by the positions of singularities of their partial wave amplitudes in the complex angular momentum plane. Already an account for leading pole singularities, so called Regge poles, allows to construct phenomenologically successful models. In particular, to explain the experimentally observed asymptotic rise of total cross-section at high energies the Regge pole with the quantum numbers of the vacuum and even parity - the Pomeron was introduced. Later it was realized that in relativistic theory Regge poles should be supplemented with Regge cuts, which could be understood as coming from the exchanges of two or more Regge poles. Next, following the pioneering work of Gribov \cite{GribovReggeonDiagrams} the reggeon field theory describing interactions between various reggeons and physical particles was developed \cite{GribovReggeonDiagrams,ReggeonFieldTheory,ReggeonCalculus}. Subsequent development of these ideas within the context of quantum chromodynamics, in particular resummation of leading high energy logarithms $(\alpha_s\ln s)^n$ to all orders in strong coupling constant (LLA resummation) with the help of Balitsky-Fadin-Kuraev-Lipatov (BFKL) equation \cite{BFKL1,BFKL2,BFKL3,BFKL4,BFKL5}, showed the LLA reggeization of QCD scattering amplitudes. The corresponding Regge pole was identified with reggeized gluon. The latter has quantum numbers of the ordinary gluon and Regge trajectory $j(t)$ passing through unity at $t=0$. Today BFKL equation is known at next-to-leading-logarithmic-approximation (NLLA) \cite{NLOBFKL1,NLOBFKL2} and the reggeization of QCD amplitudes is also proven at NLLA \cite{NLOreggeization}. In general, the amplitudes with reggeized gluons (also known as gauge invariant off-shell amplitudes) \cite{vanHamerenBCFW1,vanHamerenBCFW2,vanHamerenBCFW3,KotkoWilsonLines,vanHamerenWL1,vanHamerenWL2} arise either in the study of multi-Regge kinematics \cite{LipatovEL1,LipatovEL2,KirschnerEL1,KirschnerEL2} or
within the context of $k_T$  or high-energy factorization
\cite{GribovLevinRyskin,CataniCiafaloniHautmann,CollinsEllis,CataniHautmann}.

On the other hand, recently Roiban, Spradlin and Volovich (RSV) based on Witten's twistor string theory \cite{WittenTwistorStringTheory} got the description of N$^{k-2}$MHV$_n$ $\mathcal{N}=4$ SYM tree level amplitudes
in terms of integrals over the moduli space of degree $k-1$ curves in super twistor space  \cite{RSV,SpradlinVolovichFromTwistorString}. Subsequent generalization of RSV result by Cachazo, He and Yuan (CHY)  led to the discovery of so called {\it scattering equations} \cite{scatteringEq1,scatteringEq2,scatteringEq3,scatteringEq4,scatteringEq5}.  Within the latter tree level  $\mathcal{N}=4$ SYM amplitudes are written in terms of integrals (localized on the solutions of mentioned scattering equations) over the marked points on the Riemann sphere.  Subsequently the CHY formulae together with their loop level generalization (see \cite{TwoLoopRiemannSphere} and references therein) were derived from {\it ambitwistor string theory} \cite{AmbitwistorStringsScatEqns,ambitwistorString4d}.

Another very close research direction is related to the representation of N$^{k-2}$MHV$_n$ scattering amplitudes in terms of integrals over Grassmannians
\cite{DualitySMatrix,AmplitudesPositiveGrassmannian,AllLoopIntegrandN4SYM,GrassmanianOriginDualConformalInvariance,UnificationResidues,DualSupercondormalInvarianceMomentumTwistorsGrassmannians}. The latter naturally unifies  different BCFW \cite{BCFW1,BCFW2} representations both for tree level amplitudes and loop level integrands  \cite{DualitySMatrix,AmplitudesPositiveGrassmannian} and is ultimately connected to the integrable structure behind $\mathcal{N}=4$ SYM  S-matrix \cite{DrummondYangianSymmetry,DrummondSuperconformalSymmetry,YangBaxterScatteringAmplitudes,BetheAnsatzYangianInvariants,BeisertYangianSymmetry}.

The use of mentioned representations (Grassmannian, RSV, scattering equations and so on) of N$^{k-2}$MHV$_n$ amplitudes (i.e. the full tree level S-matrix of $\mathcal{N}=4$ SYM) provides us with relatively compact analytical expressions for $n$-point tree level amplitudes, which in their turn could be used to compute corresponding loop level amplitudes via modern unitarity based methods both at high orders of perturbation theory and/or with large number of external particles in $\mathcal{N}=4$ SYM and other field theories including QCD (see for a review \cite{Henrietta_Amplitudes}). It is important to note that these results
was almost impossible to obtain by standard Feynman diagram methods. 

The aim of the present work is to extend the recently obtained results for the usual $\mathcal{N}=4$ SYM scattering amplitudes and to derive scattering equations representation for reggeon amplitudes in $\mathcal{N}=4$ SYM from four dimensional ambitwistor string theory. At a moment, there are already scattering equations representations for the form factors of operators from stress-tensor operator supermultiplet and  scalar operators of the form $\text{Tr}(\phi^m)$ \cite{BrandhuberConnectedPrescription,HeConnectedFormulaFormFactors}. Also some formulae were extended to Standard Model amplitudes \cite{HeConnectedFormulaSM}. Besides, there are several results for the Grassmannian integral representation of form factors of operators from stress-tensor operator supermultiplet \cite{SoftTheoremsFormFactors,FormFactorsGrassmanians,WilhelmThesis,q2zeroFormFactors} and reggeon amplitudes (form factors\footnote{See the precise definition in section \ref{ReggeonsWilsonLines}.} of Wilson line insertions) \cite{offshell-1leg,offshell-multiplelegs}, see also \cite{WilsonLoopFormFactors} for a recent interesting duality for Wilson loop form factors. A very close research direction is the twistor and Lorentz harmonic chiral superspace formulation of form factors and correlation functions developed in \cite{TwistorFormFactors1,TwistorFormFactors2,TwistorFormFactors3,LHC1,LHC2,LHC3,LHC4}, see the discussion in conclusion.

This paper is organized as follows. First, in section \ref{ReggeonsWilsonLines} we introduce necessary definitions for reggeon amplitudes (Wilson lines form factors). Next, in section \ref{AmbitwistorStringsCorrelationFuncsSec} after
recalling some of the basic facts of four dimensional ambitwistor string theory we proceed with the construction of string vertex operator for reggeized gluon and derive scattering equations representation for reggeon amplitudes from corresponding string correlation functions. Finally, in section \ref{ConclusionSec} we come with our conclusion. 

\section{Reggeon amplitudes and Wilson lines}\label{ReggeonsWilsonLines}

To describe amplitudes with reggeized gluons it is convenient to use the representation of the latter in terms of Wilson line operators as in  \cite{KotkoWilsonLines}:
\begin{eqnarray}\label{WilsonLineOperDef}
\mathcal{W}_p^c(k) = \int d^4 x e^{ix\cdot k} \mathrm{Tr} \left\{
\frac{1}{\pi g} t^c \; \mathcal{P} \exp\left[\frac{ig}{\sqrt{2}}\int_{-\infty}^{\infty}
ds \; p\cdot A_b (x+ sp) t^b\right]
\right\}.
\end{eqnarray}
Here $t^c$ is $SU(N_c)$ generator\footnote{The color generators are normalized as $\mathrm{Tr} (t^a t^b) = \delta^{a b}$}, $k$ ($k^2 \neq 0$) is the reggeized gluon momentum and $p$ is its direction or polarization vector, such that $p^2=0$, $p\cdot k = 0$. The momentum and polarization vector of the reggeized gluon could be related to each other through so called  $k_T$ - decomposition of  momentum $k$:
\begin{eqnarray}\label{kT}
k^{\mu} = x p^{\mu} + k_T^{\mu} \, ,\quad x \in [0,1] \, . 
\end{eqnarray}
Note, that such decomposition could be also parametrized by an auxiliary light-cone four-vector $q^{\mu}$, so that
\begin{eqnarray}
k_T^{\mu} (q) = k^{\mu} - x(q) p^{\mu}\quad \text{with}\quad x(q) = \frac{q\cdot k}{q\cdot p} \;\; \text{and} \;\; q^2 = 0.
\end{eqnarray}
Using the fact, that the transverse momentum $k_T^{\mu}$ is orthogonal to both $p^\mu$ and $q^\mu$ vectors one can decompose it into the basis of two ``polarization'' vectors\footnote{Here we used the helicity spinor decomposition of light-like four-vectors $p$ and $q$.} \cite{vanHamerenBCFW1}:
\begin{eqnarray}
k_T^{\mu} (q) = -\frac{\kappa}{2}\frac{\la p|\gamma^{\mu}|q]}{[pq]}
- \frac{\kappa^{*}}{2}\frac{\la q|\gamma^{\mu}|p]}{\la qp\ra}\quad
\text{with} \quad \kappa = \frac{\la q|\slashed{k}|p]}{\la qp\ra},\;
\kappa^{*} = \frac{\la p|\slashed{k}|q]}{[pq]}.
\end{eqnarray}
It is easy to check, that $k^2 = -\kappa\kappa^{*}$ and both $\kappa$ and $\kappa^{*}$ variables are independent of auxiliary four-vector $q^{\mu}$ \cite{vanHamerenBCFW1}. 

Both usual and color ordered\footnote{Here we are dealing with color ordered amplitudes for simplicity. The usual amplitudes are then obtained using color decomposition, see for example \cite{offshell-1leg,DixonReview}.} reggeon amplitudes with $n$ reggeized  and  $m$ usual on-shell gluons could be then written in terms of form factors with multiple Wilson line insertions as \cite{KotkoWilsonLines}:
%
%
%
%
\begin{eqnarray}\label{AmplitudeSeveralOffShellGluons}
	\mathcal{A}_{m+n} \left(1^{\pm},\ldots ,m^{\pm},g_{m+1}^*,\ldots ,g_{n+m}^*\right) =
	\la \{k_i, \epsilon_i, c_i\}_{i=1}^m |\prod_{j=1}^n\mathcal{W}_{p_{m+j}}^{c_{m+j}}(k_{m+j})|0\ra,
\end{eqnarray}
here asterisk denotes an off-shell gluon and $p$, $k$, $c$ are its direction, momentum and color index. Next $\la \{k_i, \epsilon_i, c_i\}_{i=1}^m|=\bigotimes_{i=1}^m\la k_i,\varepsilon_i, c_i|$ and $\la k_i,\varepsilon_i, c_i|$ denotes on-shell gluon state  with momentum $k_i$, polarization vector $\varepsilon_i^-$ or $\varepsilon_i^+$ and color index $c_i$,
$p_i$ is the direction of the $i$'th ($i=1,...,n$) off-shell gluon and $k_i$ is its off-shell momentum. 
For the case when only reggeized gluons are present
(correlation function of Wilson line operators) we have:
%
%
%
\begin{eqnarray}\label{CorrFunctionOffShellGluons}
\mathcal{A}_{0+n} \left(g_1^*\ldots g_n^*\right) =
\la 0|\mathcal{W}_{p_{1}}^{c_{1}}(k_{1})\ldots \mathcal{W}_{p_{n}}^{c_{n}}(k_{n})|0\ra.
\end{eqnarray}
In the case of $\mathcal{N}=4$ SYM we may also consider other on-shell states from $\mathcal{N}=4$ supermultiplet. The most convenient way to do so is to consider colour ordered superamplitudes defined on  $\mathcal{N}=4$ on-shell momentum superspace:
%
%
%
\begin{eqnarray}\label{AmplitudeSeveralOffShellGluonsSUSY}
	A_{m+n}^* \left(\Omega_1,\ldots,\Omega_m,g_{m+1}^*,\ldots ,g_{n+m}^*\right) =
	\la \Omega_1\ldots\Omega_m|
\prod_{j=1}^n\mathcal{W}_{p_{m+j}}(k_{m+j})|0\ra,
\end{eqnarray}
where $\la \Omega_1\Omega_2\ldots\Omega_m|\equiv \bigotimes_{i=1}^m\la 0|\Omega_i$ and $\Omega_i$ ($i=1,...,m$) denotes $\mathcal{N}=4$ on-shell chiral superfield \cite{Nair}:
\begin{eqnarray}
\Omega= g^+ + \vlet_A\psi^A + \frac{1}{2!}\vlet_{A}\vlet_{B}\phi^{AB}
+ \frac{1}{3!}\vlet_A\vlet_B\vlet_C\epsilon^{ABCD}\bar{\psi}_{D}
+ \frac{1}{4!}\vlet_A\vlet_B\vlet_C\vlet_D\epsilon^{ABCD}g^{-},
\end{eqnarray}
Here, $g^+, g^-$ denote creation/annihilation operators of gluons with $+1$ and $-1$ hecilities, $\psi^A$, $\bar{\psi}_A$ stand for creation/annihilation operators of four Weyl spinors with negative helicity $-1/2$ and four Weyl spinors with  positive helicity correspondingly, while $\phi^{AB}$ denote creation/annihilation operators for six scalars (anti-symmetric in the $SU(4)_R$ $R$-symmetry indices $AB$). All $\mathcal{N}=4$ SYM fields transform  in the adjoint representation of $SU(N_c)$ gauge group.
The $A_{m+n}^* \left(\Omega_1,\ldots,g_{n+m}^*\right)$ superamplitude is then the function of the following kinematic\footnote{We used helicity spinor decomposition of on-shell particles momenta.} and Grassmann variables 
%
%
%
\begin{eqnarray}\label{AmplitudeSeveralOffShellGluonsArgumentsSUSY}
A_{m+n}^* \left(\Omega_1,\ldots,g_{m+n}^*\right) =A_{m+n}^*\left(\{\lambda_i,\tilde{\lambda}_i,\tilde{\eta}_i\}_{i=1}^{m};
\{k_i,\lambda_{p,i},\tilde{\lambda}_{p,i}\}_{i=m+1}^{m+n}\right).
\end{eqnarray}
and encodes in addition to amplitudes with gluons also  amplitudes with other on-shell states similar to the case of usual on-shell superamplitudes \cite{Henrietta_Amplitudes}.

\section{Ambitwistor string correlation functions}\label{AmbitwistorStringsCorrelationFuncsSec}

As we already mentioned in Introduction our aim here is to derive scattering equations representation for reggeon amplitudes in $\mathcal{N}=4$ SYM using four dimensional ambitwistor string theory  \cite{ambitwistorString4d}, see also  \cite{GeyerThesis} for further details. The target space of the latter is given by projective ambitwistor space $\PA$:
\begin{equation}
\PA=\bigslant{\big\{(Z,W)\in \T\times \T^* |\, Z\cdot W=0\big\}}{\{Z\cdot\partial_Z-W\cdot \partial_W\}} \, ,
\end{equation}
with $\T$ and $\T^*$ denoting twistor and dual twistor spaces. Next,  $Z = (\vll_{\alpha}, \mu^{\dotalpha}, \chi^r)\in\T = \C^{4|4}$,  $W=(\tilde \mu ,\tilde \lambda, \tilde \chi)\in\T^*$ and $Z\cdot W =\lambda_\alpha\tilde\mu^\alpha+\mu^{\dot\alpha}\tilde\lambda_{\dot\alpha}+\chi^r\tilde\chi_r$  ($r = 1,\ldots ,4$). The corresponding worldsheet theory consists from 
the worldsheet spinors $(Z, W)$ taking values in $\T\times\T^*$ together with $\text{GL}(1, \C)$ gauge field $a$ serving as a Lagrange multiplier for the constraint $Z\cdot W = 0$. In the conformal gauge its action is given by
\begin{equation}
S=\frac1{2\pi}\int_\Sigma  W\cdot \dbar Z-Z\cdot \dbar W  +a Z\cdot W +S_J\, , \label{stringAction}
\end{equation}
where $\dbar = \d\bar{\sigma}\partial_{\bar{\sigma}}$ ($\sigma, \bar{\sigma}$ are some local holomorphic and anti-holomorphic coordinates on Riemann surface $\Sigma$) and $S_J$ denotes the action for the $ \mathfrak{su}(N_c)$ worldsheet Kac-Moody current algebra  $J\in\Omega^0(\Sigma, K_\Sigma\otimes  \mathfrak{su}(N_c))$. $K_{\Sigma}$, as usual, is the canonical bundle on the surface $\Sigma$ and the other worldsheet fields take values in\footnote{Powers of the canonical bundle denote corresponding field conformal weights}
\begin{align}
&Z\in\Omega^0(\Sigma, K_\Sigma^{1/2}\otimes\T)\,,\\
&W\in\Omega^0(\Sigma,K_\Sigma^{1/2}\otimes\T^*)\,,\\
&a\in\Omega^{0,1}(\Sigma)\,,
\end{align}
To calculate string scattering amplitudes we need vertex operators. There are two equivalent representations for integrated vertex operator used to describe on-shell states \cite{ambitwistorString4d}:
\begin{align}
&\mathcal{V}_a=\int\frac{\d s_{a}}{s_{a}}\bar{\delta}^{2}(\lambda_{a}-s_{a}\lambda)\e^{is_{a}\left([\mu\,\tilde{\lambda}_{a}]+\chi^{r}\tilde{\eta}_{ar}\right)} J\cdot T_a \,, \\
&\widetilde{\mathcal{V}}_a= \int\frac{\d s_{a}}{s_{a}}\bar{\delta}^{2|4}(\tilde{\lambda}_{a}-s_{a}\tilde{\lambda}
|\tilde{\eta}_a-s_a\tilde{\chi} )\e^{is_{a}\la\tilde{\mu}\,\lambda_{a}\ra}J\cdot T_a\, ,
\end{align}
where $\bar{\delta}(z) = \dbar (1/2\pi i z)$. We would like to stress here, that both vertex operators contain all sixteen on-shell states of $\mathcal{N}=4$ SYM. To obtain $\text{N}^{k-2}\text{MHV}$ on-shell scattering amplitudes one may use for example the correlation function of $k$ $\widetilde{\mathcal{V}}$ operators and $n-k$ $\mathcal{V}$ operators \cite{ambitwistorString4d}:
\begin{equation}
A_{k,n}=\left\la \widetilde{\mathcal{V}}_1 \ldots \widetilde{\mathcal{V}}_k \mathcal{V}_{k+1}\ldots  \mathcal{V}_n\right\ra\, .
\end{equation}
This correlation function is not hard to calculate\footnote{See \cite{ambitwistorString4d} for details} and we get\cite{ambitwistorString4d}:
\begin{equation}
\begin{split}
A_{n,k}=\int  \frac{1}{\text{Vol} \,\text{GL}(2,\C)}
&\prod_{a=1}^n\frac{\d s_a\d\sigma_a}{s_a (\sigma_a-\sigma_{ a+1})}
\prod_{p=k+1}^n\bar\delta^{2}
(\lambda_p-s_p\lambda(\sigma_p)) \\
&\prod_{i=1}^k  \bar\delta^{2|4}(\tilde \lambda_i -s_i\tilde\lambda (\sigma_i), \tilde{\eta}_i-s_i\tilde{\chi}(\sigma_i)) \, ,
\end{split} \label{onshellStringAmp}
\end{equation}
where
\begin{align}
&\lambda (\sigma)=\sum_{i=1}^{k}\frac{s_{i}\lambda_{i}}{\sigma-\sigma_{i}}\,,\quad \tilde{\lambda} (\sigma) = \sum_{p=k+1}^{n}\frac{s_{p} \tilde{\lambda}_{p}}{\sigma-\sigma_{p}}\, , \quad \tilde{\chi} (\sigma) = \sum_{p=k+1}^{n}\frac{s_{p} \tilde{\eta}_p}{\sigma-\sigma_{p}}\, .  \label{ZWsolution1}
\end{align}
In terms of homogeneous coordinates on Riemann sphere $\sigma_\alpha=\frac1s (1,\sigma)$ the same result is written as \cite{ambitwistorString4d}:
\begin{equation}
\begin{split}
A_{n,k}=\int  \frac1{\text{Vol} \,\text{GL} (2,\C)} &\prod_{a=1}^n\frac {\d^2 \sigma_a}{(a\, a+1)}  \, \prod_{p=k+1}^n\bar\delta^{2} (\lambda_p-\lambda(\sigma_p)) 
\prod_{i=1}^k  \bar\delta^{2|4}(\tilde \lambda_i - \tilde\lambda(\sigma_i),
\tilde{\eta}_i-\tilde{\chi}(\sigma_i))\, 
\end{split} \label{onshellStringAmpCovariant}
\end{equation}
with $(i \,j)=\sigma_{i\alpha}\sigma_j^\alpha$ and
\begin{equation}
\lambda (\sigma) = \sum_{i=1}^{k}\frac{\lambda_{i}}{(\sigma\, \sigma_{i})}
\, , \quad 
\tilde{\lambda} (\sigma) = \sum_{p=k+1}^{n}\frac{\tilde{\lambda}_{p}}{(\sigma \,\sigma_{p})}\, , \quad
\tilde{\chi} (\sigma) = \sum_{p=k+1}^{n}\frac{\tilde{\eta}_p}{(\sigma \,\sigma_{p})}\, .
\end{equation}
The scattering equations are then straightforwardly follow from the arguments of delta functions:
\begin{equation}
k_a\cdot P(\sigma_a)=\lambda_a^\alpha\tilde\lambda_a^{\dot\alpha} P_{\alpha\dot\alpha}(\sigma_a)=\lambda_a^\alpha\tilde\lambda_a^{\dot\alpha}  \lambda_\alpha(\sigma_a)\tilde\lambda_{\dot\alpha}(\sigma_a) = 0\,.
\end{equation}

To describe reggeon amplitudes we also need the ambitwistor string vertex operator for reggeized gluon. The latter could be obtained from the pullback of corresponding ambitwistor space wave functions to string theory worldsheet\footnote{For the previous work on reggeon string vertexes within superstring theory see \cite{StringRegge1,StringRegge2} and references therein.}. The required ambitwistor space wave function could be easily found using a representation of corresponding reggeon amplitudes with $n+1$ legs in terms of convolutions of particle-particle-reggeon PPR vertexes (minimal off-shell amplitudes in the language of \cite{offshell-1leg,offshell-multiplelegs}) with on-shell amplitudes with $n+2$ legs. This construction comes naturally by noting that gluing (introducing loop integration) PPR vertex to the on-shell amplitude we get one-loop reggeon amplitude whose leading singularity (extracted by maximally cutting the loop) gives us the corresponding tree level reggeon amplitude. Thus, the ambitwistor string vertex operator for reggeized gluon (Wilson line operator insertion) may be written as
 \begin{equation}
 {\mathcal V}^{\text{WL}}_{n, n+1} = \int\prod_{i=n}^{n+1}
 \frac{\d^2\vll_i\d^2\vlt_i}{\text{Vol[GL(1)]}} \d^4\vlet_i A^*_{2,2+1} (\Omega_n, \Omega_{n+1}, g^*) \Big|_{\vll\to -\vll} {\mathcal V}_{n} \mathcal{V}_{n+1} \Big|_{T^a T^b\to i f^{abc} T^c\to T^c} \, ,
 \end{equation}
Here, $c$ denotes the color index of the reggeized gluon and we have used projection of tensor product of two adjoint on-shell gluon color representations onto reggeized gluon adjoint color representation. The minimal PPR vertex $A^*_{2,2+1} (\Omega_n, \Omega_{n+1}, g^*)$ is given by \cite{offshell-1leg}:
%
%
%
%
%
%
%
\begin{eqnarray}
 A_{2,2+1}^*(\Omega_n, \Omega_{n+1}, g^*) = \frac{\delta^4(k+\vll_{n}\vlt_{n}+\vll_{n+1}\vlt_{n+1})}{\kappa^*}\frac{\delta^4\left(\tilde{\eta}_{n}\abr{p\, n+1}+\tilde{\eta}_{n+1}\abr{p\, n}\right)}
 {\abr{p\, n}\abr{n\, n+1}\abr{n+1\, p}}\, ,
 \end{eqnarray}
 where $p=\lambda_p\tilde{\lambda}_p$ is the reggeized gluon direction and $\kappa^*$ was defined in Section \ref{ReggeonsWilsonLines} when introducing $k_T$ decomposition of the reggeized gluon momentum $k$. It should be noted, that each of $\mathcal{V}$ vertex operators above could be exchanged for $\widetilde{\mathcal V}$ operator and thus the above representation for reggeized gluon vertex operator is not unique. The ambitwistor string vertex operator we got is non-local by construction. The latter property is expected as Wilson line is non-local object by itself. Performing integrations\footnote{See \cite{FormFactorsGrassmanians,offshell-1leg} for details.} over helicity spinors $\vll_i, \vlt_i$ we get (the projection operator $\partial^4_{\eta_{p}}$ acting on ${\mathcal V}_{n} \mathcal{V}_{n+1}$ is assumed)
 \begin{equation}
  {\mathcal V}^{\text{WL}}_{n, n+1} = \frac{\la\xi p\ra}{\kappa^*}
  \int\frac{\d\beta_2}{\beta_2}\int\frac{\d\beta_1}{\beta_1}\frac{1}{\beta_1^2\beta_2}
  {\mathcal V}_{n} \mathcal{V}_{n+1} \Big|_{T^a T^b\to i f^{abc} T^c\to T^c} \, ,
 \end{equation}
 where
\begin{align}\label{spinorsAsFunctionsOfBeta}
 &\vll_{n} = \vlluu_{n} + \beta_2\vlluu_{n+1}\, , && \vlt_{n} =
 \beta_1\vltuu_{n}  + \frac{(1+\beta_1)}{\beta_2}\vltuu_{n+1}\, ,
 &&\vlet_{n} = -\beta_1\vleuu_{n}\, , \\
&\vll_{n+1} = \vlluu_{n+1} + \frac{(1+\beta_1)}{\beta_1\beta_2}\vlluu_{n}\, ,
&& \vlt_{n+1} = -\beta_1\vltuu_{n+1} -\beta_1\beta_2 \vlluu_{n}\, , &&\vlet_{n+1} = \beta_1\beta_2\vleuu_{n}\, .
 \end{align}
with
\begin{eqnarray}\label{kDecompsition}
    \vlluu_{n}=\lambda_p,~\vltuu_{n}=\frac{\la\xi |k}{\la\xi p\ra},~\vleuu_{n}=\vlet_p;
    ~~\vlluu_{n+1}=\lambda_{\xi},~\vltuu_{n+1}=\frac{\la p |k}{\la\xi p\ra},~\vleuu_{n+1}=0,
\end{eqnarray}
where $\lambda_{\xi} \equiv\la\xi |$ is some arbitrary spinor. In practical calculations it is useful to identify it with the spinor $\lambda_{q}$ coming from helicity spinor decomposition of auxiliary momentum $q$ arising in $k_T$ decomposition of reggeized gluon momentum $k$.

The reggeon amplitude with one reggeized gluon and $n$ on-shell final states is then given by the following ambitwistor string correlation function:
\begin{equation}
A^*_{k,n+1} =
\left\la \widetilde{\mathcal{V}}_1 \ldots \widetilde{\mathcal{V}}_k \mathcal{V}_{k+1}\ldots  \mathcal{V}_n {\mathcal V}^{\text{WL}}_{n+1, n+2}\right\ra\, .
\end{equation}
Evaluating first string correlator of on-shell vertexes with the help of (\ref{onshellStringAmp}) we get
\begin{align}
A^*_{k,n+1} = &\frac{\la\xi p\ra}{\kappa^*}
\int\frac{\d\beta_2}{\beta_2}\int\frac{\d\beta_1}{\beta_1}\frac{1}{\beta_1^2\beta_2}\, \frac{1}{\text{Vol} \,\text{GL} (2,\C)} \nonumber \\ &\times
\int \prod_{a=1}^{n+2}\frac{\d s_a\d\sigma_a}{s_a (\sigma_a-\sigma_{ a+1})}
\prod_{p=k+1}^{n+2}\bar\delta^{2}
(\lambda_p-s_p\lambda(\sigma_p))
\prod_{i=1}^k  \bar\delta^{2|4}(\tilde \lambda_i -s_i\tilde\lambda (\sigma_i), \tilde{\eta}_i-s_i\tilde{\chi}(\sigma_i)) \, . \label{offshellamp1}
\end{align}
Now using unity decomposition as in  \cite{UnificationResidues}:
\begin{equation}\label{ResolitionOfIndentity}
1 = \frac{1}{\text{Vol} \,\text{GL} (k)}\int\d^{k\times (n+2)}C\, \d^{k\times k} L\, (\det L)^{n+2}
\delta^{k\times (n+2)} \left(C - L\cdot C^V [s, \sigma]\right)\, ,
\end{equation}
where the integral over $L$ matrix is the integral over $\text{GL} (k)$ transformations and $C^V [\sigma]$ denotes the Veronese map from $(\C^2)^{n+2}/\text{GL} (2)$ to $G(k, n+2)$ Grassmannian \cite{UnificationResidues} (see also \cite{BrandhuberConnectedPrescription}):
\begin{align}\label{VeroneseMap}
C^V [s, \sigma] = \left(
\begin{array}{cccc}
\vdots & \vdots & \cdots & \vdots \\
\sigma^V [s_1, \sigma_1] & \sigma^V [s_2, \sigma_2] & \cdots & \sigma^V [s_{n+2}, \sigma_{n+2}] \\
\vdots & \vdots & \cdots & \vdots \end{array}
\right)\, ,\quad  \sigma^V [s, \sigma]\equiv \left(\begin{array}{c}
\xi \\ \xi\sigma \\ \vdots \\ \xi\sigma^{k-1}
\end{array}\right)\, ,
\end{align}
where \cite{SpradlinVolovichFromTwistorString,BrandhuberConnectedPrescription} :
\begin{align}
& \xi_i = s_i^{-1}\prod_{j=1, j\neq i}^k (\sigma_j - \sigma_i)^{-1}\, , && i\in (1,k) \\
& \xi_i = s_i\prod_{j=1}^k (\sigma_j - \sigma_i)^{-1}\, , && i\in (k+1,n+2)
\end{align}
Next, integrating (\ref{offshellamp1}) over  $s_a$ and $\sigma_a$ we get
\begin{align}
A^*_{k,n+1} = &\frac{\la\xi p\ra}{\kappa^*}
\int\frac{\d\beta_2}{\beta_2}\int\frac{\d\beta_1}{\beta_1}\frac{1}{\beta_1^2\beta_2}\, \frac{1}{\text{Vol} \,\text{GL} (k)} \nonumber \\ &\times
\int\d^{k\times (n+2)} C\, F (C)\, \delta^{k\times 2} (C\cdot\vlt)\delta^{k\times4} (C\cdot \vlet) \delta^{(n+2-k)\times 2} (C^{\perp}\cdot\vll)\, ,  \label{offshellamp2}
\end{align}
where
\begin{align}
F(C) = \int \frac{1}{\text{Vol} \,\text{GL} (2,\C)}\prod_{a=1}^{n+2}\frac{\d s_a\d\sigma_a}{s_a (\sigma_a-\sigma_{ a+1})} \d^{k\times k} L\,
\delta^{k\times (n+2)} \left(C - L\cdot C^V [s, \sigma]\right),
\end{align}
and
\begin{eqnarray}
    \delta^{k\times 2} (C\cdot\vlt)&\equiv&\prod_{a=1}^k\delta^{2} \left(\sum_{i=1}^n C_{ai}\vlt_i\right),~\delta^{(n+2-k)\times 2} (C^{\perp}\cdot\vll)\equiv\prod_{b=k+1}^{n+2}
    \delta^{2} \left(\sum_{j=1}^{n+2}C^{\perp}_{bj}\vll\right),\nonumber\\
    \delta^{k\times 4} (C\cdot \vlet)&\equiv&\prod_{a=1}^k\hat{\delta}^{4} \left(\sum_{i=1}^n C_{ai}\vlet_i\right),
\end{eqnarray}
Here, $C^{\perp}$ is the matrix defined by the identity $C\cdot (C^{\perp})^T=0$ and it is assumed that all matrix manipulations are performed after $GL(k)$ gauge fixing. Also note that the above delta functions should be thought as $\delta (x) = 1/x$ and the corresponding contour integral will then compute the residue at $x =0$ \cite{ArkaniHamedPhysicsFromGrassmannian}. 

By construction $F (C)$ contains $(k-2)\times (n-k)$ delta function factors  forcing integral over $C$ matrix to have Veronese form \cite{UnificationResidues}. In general $F(C)$ is a rather complicated rational function of the minors of $C$ matrix. 
For example,
for $k=3$ and $n+2=6$ it is given by \cite{UnificationResidues,VolovichAGrassmannianEtudeinNMHVMinors}:
\begin{eqnarray}
    F(C)&=&\frac{(135)}{(123)(345)(561)}\frac{1}{S},\nonumber\\
    S&=&(123)(345)(561)(246)-(234)(456)(612)(351).
\end{eqnarray}
Here\footnote{We hope there will be no confusion with previous definition $(i \,j)=\sigma_{i\alpha}\sigma_j^\alpha$ used in $d^2\sigma_a$ integrals over homogeneous coordinates on Riemann sphere before.}  $(i_1\ldots i_k)$ is minor of $C$ matrix constructed from $i_1,\ldots, i_k$ columns of $C$. The integral over $d^{k\times (n+2)} C/\text{Vol} \,\text{GL} (k,\C)$ can then be reduced
to multidimensional contour integral over $(k-2)(n-k)$ complex variables $\tau$ and evaluated by taking residues.
However, at the end after all technicalities (which are highly non-trivial and interesting in their own turn \cite{UnificationResidues,VolovichAGrassmannianEtudeinNMHVMinors}) it could be shown that $F(C)$ may be chosen in the form 
\begin{equation}
F(C) = \frac{1}{(1\cdots k) (2\cdots k+1)\cdots (n+2\cdots k-1)}.
\end{equation}
Next, let us rewrite (\ref{offshellamp2}) in the form (the proper choice of integration contour $\Gamma_{k,n+2}^{tree}$ is implemented \cite{UnificationResidues,GoddardGluonTreeAmplitudesinOpenTwistorString}) 
\begin{multline}
A^{*}_{k,n+1} = \frac{\la\xi p\ra}{\kappa^{*}} \int
\frac{d^{k\times (n+2)} C}{\text{Vol}[GL(k)]}
\frac{d\beta_1 d\beta_2}{\beta_1\beta_2^2}\frac{\delta^{k\times 2} \left( C'
	\cdot \vltuu \right)
	\delta^{k\times 4} \left(C'\cdot \vleuu \right)
	\delta^{(n+2-k)\times 2} \left(C'^{\perp}\cdot \vlluu \right)}{(1\cdots k)(2\cdots k+1)\cdots (n+2\cdots k-1)} ,
\end{multline}
where 
\begin{align}
C'_{n+1} &= -\beta_1 C_{n+1} + \beta_1\beta_2 C_{n+2},
&C'^{\perp}_{n+1} &= C^{\perp}_{n+1} + \frac{1+\beta_1}{\beta_1\beta_2} C^{\perp}_{n+2}, \nonumber \\
C'_{n+2} &= -\beta_1 C_{n+2} + \frac{1+\beta_1}{\beta_2} C_{n+1},
&C'^{\perp}_{n+2} &= C^{\perp}_{n+2} + \beta_2 C^{\perp}_{n+1},
\end{align}
and
\begin{align}\label{SpinorsInDeformadGrassmannian2}
&\vlluu_i = \vll_i, & i = 1,\ldots  n& , &\vlluu_{n+1} &= \lambda_p ,
&\vlluu_{n+2} &= \xi \nonumber \\
&\vltuu_i = \vlt_i, & i = 1,\ldots  n& ,
&\vltuu_{n+1} &= \frac{\la\xi |k}{\la\xi p\ra},
&\vltuu_{n+2} &= - \frac{\la p |k}{\la\xi p\ra} , \nonumber \\
&\vleuu_i = \vlet_i,  & i = 1,\ldots  n& , &\vleuu_{n+1} &= \vlet_p ,
&\vleuu_{n+2} &= 0 . \nonumber \\
\end{align}
Now, introducing inverse $C$-matrix transformation
\begin{align}
C_{n+1} &= C'_{n+1} +\beta_2 C'_{n+2}, \nonumber \\
C_{n+2} &= \frac{1+\beta_1}{\beta_1\beta_2} C'_{n+1} + C'_{n+2} ,
\end{align}
minors of $C$-matrix containing both $n+1$ and $n+2$ columns when rewritten in terms of minors of $C'$-matrix acquire extra $-\frac{1}{\beta_1}$ factor. For example, for $(n+1\cdots k-2)$ minor we have
\begin{align}
(n+1\cdots k-2) = -\frac{1}{\beta_1} (n+1\cdots k-2)'\, .
\end{align}
On the other hand, minors containing either $n+1$ or $n+2$ column transform as
\begin{align}
(n+2\, 1\cdots k-1) &= \frac{1+\beta_1}{\beta_1\beta_2}(n+1\, 1\cdots k-1)' + (n+2\, 1\cdots k-1)'\, , \\
(n-k+2\cdots n+1) &= (n-k+2\cdots n+1)' + \beta_2 (n-k+2\cdots n\, n+2)'\, ,
\end{align}
while all other minors remain unchanged $(\cdots) = (\cdots)'$. Now, going to the integral over $C'$ matrix and accounting for the transition Jacobian $\left(-\frac{1}{\beta_1}\right)^k$  we get
\begin{multline}
A^{*}_{k,n+1} = -\frac{\la\xi p\ra}{\kappa^{*}} \int
\frac{d^{k\times (n+2)} C'}{\text{Vol}[GL(k)]}
\frac{d\beta_1 d\beta_2}{\beta_1\beta_2}\delta^{k\times 2} \left( C'
\cdot \vltuu \right)
\delta^{k\times 4} \left(C'\cdot \vleuu \right)
\delta^{(n+2-k)\times 2} \left(C'^{\perp}\cdot \vlluu \right) \nonumber \\
\times \frac{1}{(1\cdots k)'\cdots (n+2\cdots k-1)'\left(1+\beta_2\frac{(n-k+2\cdots n\, n+2)'}{(n-k+2\cdots n\, n+1)'}\right)\left(\beta_1\beta_2 + (1+\beta_1)\frac{(n+1\, 1\cdots k-1)'}{(n+2\, 1\cdots k-1)'}\right)}	
\end{multline}
Next, taking first residue at $\beta_2 = 0$ and then at $\beta_1 = -1$ (i.e. considering corresponding residual form) we recover our previous result for reggeon amplitude with one reggeized gluon\cite{offshell-1leg} (here we again assume $\partial^4_{\eta_{p}}$ projection operator acting on Grassmannian integral):
\begin{equation}
A^*_{k,n+1} = \int_{\Gamma_{k,n+2}^{tree}}\frac{d^{k\times
		(n+2)}C'}{\text{Vol}[GL (k)]}Reg. \frac{\delta^{k\times 2} \left( C'
	\cdot \vltuu \right)
	\delta^{k\times 4} \left(C' \cdot \vleuu \right)
	\delta^{(n+2-k)\times 2} \left(C'^{\perp} \cdot \vlluu \right)}{(1 \cdots k)'\cdots (n+1 \cdots k-2)' (n+2 \; 1\cdots k-1)'}, \label{offshellAmpGrassmannian}
\end{equation}
with\footnote{The $Reg.$ notation is chosen because this ratio of minors regulates soft holomorphic limit with respect to external kinematical variables associated with reggeized gluon \cite{offshell-1leg}.}
\begin{eqnarray}
Reg.=\frac{\la\xi p\ra}{\kappa^{*}}\frac{(n+2 \; 1\cdots k-1)'}{(n+1 \; 1\cdots k-1)'}.
\end{eqnarray}
We  have also verified that direct evaluation of (\ref{offshellamp1}) reproduces all particular off-shell amplitudes considered as examples in \cite{offshell-1leg}.

Finally we need to perform inverse operation, that is to reduce integral in (\ref{offshellAmpGrassmannian}) to the integral over $G(2, n+2)$ Grassmannian. This is again done with the help of Veronese map \cite{UnificationResidues,BrandhuberConnectedPrescription}. Using resolution of unity (\ref{ResolitionOfIndentity}), fixing $GL(k)$ gauge to enforce rational form of scattering equations \cite{ambitwistorString4d} and performing integration over $C'$ matrix  our Grassmannian integral representation (\ref{offshellAmpGrassmannian}) takes the form of scattering equations representation we are looking for:
\begin{align}
\int
\prod_{a=1}^{n+2}\frac {\d^2 \sigma_a}{(a\, a+1)}\frac{Reg.^V}{\text{Vol} \,\text{GL} (2,\C)} \prod_{p=k+1}^{n+2}\bar\delta^{2} (\vlluu_p-\vlluu(\sigma_p))
\prod_{i=1}^k  \bar\delta^{2|4}(\vltuu_i - \vltuu(\sigma_i),
\vletuu_i-\tilde{\underline{\underline{\chi}}}(\sigma_i))\, ,
\end{align}
where
\begin{equation}
Reg.^V = \frac{\la\xi p\ra}{\kappa^*}\frac{(k\, n+1)}{(k\, n+2)}
\end{equation}
and we have also performed the transition to homogeneous coordinates on Riemann sphere. The doubly underlined functions are defined as 
\begin{equation}
\vlluu = \sum_{i=1}^{k}\frac{\vlluu_{i}}{(\sigma\, \sigma_{i})}
\, ,\quad
\vltuu = \sum_{p=k+1}^{n+2}\frac{\vltuu_{p}}{(\sigma \,\sigma_{p})}\, ,\quad
\tilde{\underline{\underline{\chi}}} = \sum_{p=k+1}^{n+2}\frac{\vleuu_p}{(\sigma \,\sigma_{p})}\, .
\end{equation}
The result for the case of reggeon amplitudes with multiple reggeized gluons $A^*_{m+n}$ could be obtained along the same lines. For example, in the case with first $m$ particles on-shell and last $n$ being reggeized gluons we would get:
\begin{align}\label{MultipleOff-shellGluonsAmplRVS}
 \int
 \prod_{a=1}^{n+2}\frac {\d^2 \sigma_a}{(a\, a+1)}\frac{Reg.^V(m+1,\ldots , m+n)}{\text{Vol} \,\text{GL} (2,\C)} \prod_{p=k+1}^{m + 2 n}\bar\delta^{2} (\vlluu_p-\vlluu(\sigma_p))
 \prod_{i=1}^k  \bar\delta^{2|4}(\vltuu_i - \vltuu(\sigma_i),
 \vletuu_i-\tilde{\underline{\underline{\chi}}}(\sigma_i))\, ,
\end{align}
where
\begin{gather}
 Reg.^V (m+1,\ldots , m+n) = \prod_{j=1}^{n} Reg^V. (j+m)\, ,\quad
 Reg.^V (j+m) = \frac{\la\xi_j p_j\ra}{\kappa_j^*}\frac{(k\, 2j-1+m)}{(k\, 2j+m)}\, .
\end{gather}
and external kinematical variables are defined as
\begin{align}\label{SpinorsInDeformadGrassmannianMultipleOffShellMomenta}
 &\vlluu_i = \vll_i, & i = 1,\ldots  m& ,
 &\vlluu_{m+2j-1} &= \lambda_{p_j} ,
 &\vlluu_{m+2j} &= \xi_j, & j = 1,\ldots  n, \nonumber \\
 &\vltuu_i = \vlt_i, & i = 1,\ldots  m& ,
 &\vltuu_{m+2j-1} &= \frac{\la\xi_j |k_{m+j}}{\la\xi_j p_j\ra},
 &\vltuu_{m+2j} &= - \frac{\la p_j |k_{m+j}}{\la\xi_j p_j\ra} , & j = 1,\ldots  n, \nonumber \\
 &\vleuu_i = \vlet_i,  & i = 1,\ldots  m& , &\vleuu_{m+2j-1} &= \vlet_{p_j} ,
 &\vleuu_{m+2j} &= 0 , & j = 1,\ldots  n. \nonumber \\
 \end{align} 
Note, that it is possible to rewrite (\ref{MultipleOff-shellGluonsAmplRVS}) as an integral over
$Gr(k,m+2n)$ Grassmannian coinciding with our previous result\cite{offshell-multiplelegs}.

\section{Conclusion}\label{ConclusionSec}

In this paper we presented results for scattering equations representations for reggeon amplitudes in $\mathcal{N}=4$ SYM derived from four dimensional ambitwistor string theory. The presented derivation could be also easily generalized to the case of tree level form factors of local operators and loop integrands of reggeon amplitudes, which will be the subject of our forthcoming publication \cite{FormFactorsAmbitwistorString}. 
 
As by product we found an easy and convenient gluing procedure allowing us to obtain required reggeon amplitude expressions from already known on-shell amplitudes. The construction of string vertex operator for reggeized gluon was inspired by the mentioned gluing procedure.  It would be extremely interesting to  consider pullbacks of composite operators defined on twistor or Lorentz harmonic chiral superspace \cite{TwistorFormFactors1,TwistorFormFactors2,TwistorFormFactors3,LHC1,LHC2,LHC3,LHC4} to construct corresponding string vertex operators. We hope that along these lines we will be able to get scattering equations representation for arbitrary local composite operators.

Having obtained scattering equations representations one may wonder what is the most efficient way to
get final expressions for amplitudes with given numbers of reggeized gluons and other on-shell states. In the case of usual on-shell amplitudes we know that they could be obtained through the computations of global residues by the methods of computational algebraic geometry \cite{GlobalResidue1,GlobalResidue2,GlobalResidue3} , see also \cite{Talesof1001Gluons}.  It would be interesting to see how this procedure works in the case of reggeon amplitudes considered here.

Finally, we should also develop methods for computing 
loop corrections to reggeon amplitudes together with their loop level generalization of scattering equations representation. Besides, it is extremely interesting to see how the presented approach works in gravity and supergravity theories, where we have a well developed framework for reggeon amplitudes based on high-energy effective lagrangian, see \cite{gravityEL4,gravityEL5} and references therein.

\section*{Acknowledgements}

The authors would like to thank D.I. Kazakov, L.N. Lipatov and  Yu-tin Huang for interesting and stimulating discussions. This work was supported by RSF grant \#16-12-10306.

\bibliographystyle{utphys2}
\bibliography{refs}

\end{document}